\documentstyle[11pt,newpasp,twoside,epsf]{article}
\def\edcomment#1{\iffalse\marginpar{\raggedright\sl#1\/}\else\relax\fi}
\begin{document}
\title{Studies of Structure Formation and Cosmology with Galaxy Cluster Surveys}
\author{Joseph J. Mohr}
\affil{Departments of Astronomy and Physics;
University of Illinois, 1002 W. Green St.; Urbana, IL  61801 USA}

\begin{abstract}

Surveys of galaxy clusters provide a promising method of testing 
models of structure formation in the universe.  Within the context of
our standard structure formation scenario, surveys provide 
measurements of the geometry of the universe and the nature of the 
dark energy and dark matter.  Cluster catalogues will be constructed 
using some combination of X--ray, optical/near--IR, and mm or cm-wave 
observations.  These catalogues will be used to study the cluster redshift and 
mass distributions along with the correlations of the cluster spatial 
distribution.  These measurements probe the volume--redshift relation, 
the power spectrum of density fluctuations and the 
evolution of galaxy cluster abundance.  All are sensitive to the amount 
of dark matter $\Omega_{M}$, the amount of dark energy $\Omega_{E}$, 
the equation of state of the dark energy $w(z)$ and any other parameter, 
which affects the expansion history of the universe.

\end{abstract}

\section{Introduction}
\label{sec:intro}

Over the last few years, cosmological constraints from Type Ia 
SNe (Schmidt et al. 1998; Perlmutter et al. 1999), cluster baryon fractions 
(White et al. 1993a; David et al. 1995; White \& Fabian 1995; 
Burles \& Tytler 1998; Mohr et al. 1999; Arnaud \& Evrard 1999), 
the cosmic microwave background (CMB) anisotropy 
(Hanany et al. 2000; Jaffe et al 2000; Lange et al. 2001) 
and other complementary measures (Bahcall et al. 1999, and references therein)
have pointed toward a dark energy dominated 
universe ($\Omega_\Lambda\sim{2\over3}$),
with a significant dark matter component ($\Omega_m\sim{1\over3}$) and a trace
of baryonic matter.  The recent detections of
the 2nd and 3rd acoustic peaks in the CMB anisotropy 
(Halverson et al. 2001; Netterfield et al. 2001; Pryke et al. 2001)
lend additional support to these conclusions and bring several
important questions into sharp focus.
At the dawn of this new era of precision cosmology, the important
questions concern the very nature of the dark matter (collisionless or 
self-interacting) and the characteristics
of the dark energy (which we can parametrize by the equation of state
parameter $w$, where the pressure $p=w\rho$).

Recent theoretical and experimental developments make future
cosmological studies that utilize galaxy clusters extremely promising.
One particularly promising approach is the use of galaxy cluster
surveys, which enable one to measure the cluster redshift 
distribution and the correlations in the cluster spatial distribution.
Surveys are now being carried out using cluster X--ray 
emission, the near-IR/optical light from cluster galaxies, the 
distorted morphologies and alignment of background galaxies, and 
the effect that hot electrons within clusters have on the 
cosmic microwave background (the so-called Sunyaev-Zel'dovich effect 
or SZE; Sunyaev \& Zel'dovich 1972).  To use these surveys to 
full effect in cosmology studies, we must first test the standard 
model of structure formation.  In addition, we must 
sharpen our understanding of the nature and evolution of 
galaxy cluster internal structure and the relationships between
cluster observables (i.e. SZE decrement, X-ray emission, galaxy light)
and the cluster halo mass.

In these proceedings we describe a fundamental test of the 
hierarchical structure formation model, and then we examine in some 
detail the cosmological dependences of the cluster redshift 
distribution.  We end by highlighting some of the challenges that 
currently exist in using cluster surveys to precisely constrain 
cosmological quantities like the equation of state of the dark energy.

\section{Structure Formation Constraints from High-$z$ Cluster Surveys}

Because of the nature of the power spectrum of density fluctuations, 
we expect that structure formation proceeded hierarchically from 
small to ever larger scales (i.e. Peebles 1993 and references therein).  
Low mass galaxy clusters 
($\sim10^{14}M_{\odot}$) are expected 
to first emerge at redshifts of $z=2$ to $z=3$ within the currently 
favored model.  Higher mass clusters ($\sim10^{15}M_{\odot}$) appear 
later at lower redshifts.
An appealingly powerful test of structure formation 
would be to probe the cluster population with sufficient sensitivity 
to detect the first emerging low mass systems.
High sensitivity SZE surveys are particularly well suited for studies 
of the high redshift galaxy cluster population, because of the 
redshift independence of the decrement $\Delta T$:
\begin{equation}
{\Delta T\over T_{cmb}} = -2 {\sigma_{T}\over m_{e}c^{2}}
\int\, dl n_{e}k_{B}T_{e},
\end{equation}
where $T_{cmb}$ is the cosmic microwave background (CMB) temperature, 
$\sigma_{T}$ is the Thomson cross section, $m_{e}$ is the electron rest 
mass, $c$ is the speed of light, $k_{B}$ is the Boltzmann constant 
and $n_{e}$ and $T_{e}$ are the 
electron number density and temperature.  In other words, if one has 
a galaxy cluster described by a particular distribution of $n_{e}$ 
and $T_{e}$, the magnitude of the SZE distortion of the CMB along a 
line of sight passing through the cluster would be independent of the 
cluster redshift.  This together with our expectation for how cluster 
structure evolves with redshift, makes SZE instruments capable of 
detecting clusters of a particular mass no matter what that cluster's 
redshift (Holder et al. 2000).  This is a particularly power approach 
for determining the redshifts when galaxy clusters first emerged.

\begin{figure}[htb]
\hbox to \hsize{\epsfysize=1.75in\epsfbox{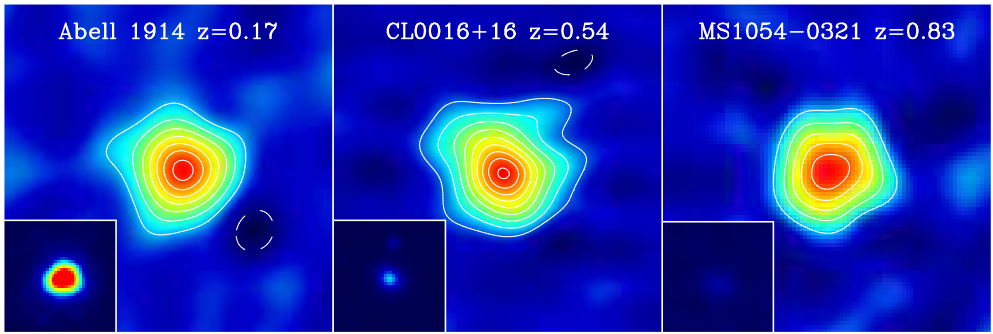}\hfil}
\caption{Interferometric SZE and ROSAT X-ray (inset) images of three galaxy 
clusters of comparable mass at redshifts $z=0.2$, 0.5 and 0.8.  SZE 
contours are 75~$\mu$K, and the X-ray color scale is the same in each 
cluster.  Note that while the SZE signal remains 
comparable at increasing redshift, the X--ray surface brightness 
suffers cosmological dimming.  This characteristic makes SZE 
observations particularly well suited to studies of high redshift 
clusters. (figure courtesy J.E. Carlstrom and J.J. Mohr)}
\end{figure}

In contrast to this SZE behavior, cluster X-ray emission (along with 
any emission) suffers from $\left(1+z\right)^{4}$ cosmological 
dimming.  The X--ray surface brightness $I_{x}$ is
\begin{equation}
I_{x} = {1\over 4\pi\left(1+z\right)^{4}}{\mu_{e}\over\mu_{H}}
\int\, dl n^{2}_{e}\Lambda(T_{e}),
\end{equation}
where $n_{e}m_{p}\mu_{e}\equiv\rho$, $m_{p}$ is the proton rest 
mass, $\rho$ is the intracluster medium mass density, and $\Lambda$ is
the temperature dependent X--ray emission coefficient describing 
bremsstrahlung and line emission.  This strikingly different behavior 
of X--ray emission and the SZE is qualitatively illustrated in 
Figure~1 by the 
panel of interferometric SZE observations (with X--ray image insets) 
of three clusters of comparable mass at redshifts $z=0.2$, 0.5 and 
0.8.  The SZE contours and X-ray color scales are the same for all 
three clusters.  Although the cluster SZE signal is similar at all 
redshifts, the X--ray emission dims rapidly, as expected.

High sensitivity interferometric SZE surveys carried out with a new 
generation of SZE optimized interferometers will soon carry out the 
fundamental test of hierarchical structure formation described 
above.  Three such instruments, the SZ-Array, AMiBA and AMI (all 
described elsewhere in this volume), are all funded and currently in 
various stages of construction.

\section{Cosmological Constraints from the Cluster Redshift Distribution}

Within the context of the standard structure formation scenario, it 
is possible to use cluster surveys to measure cosmological parameters.
The abundance of galaxy clusters and its redshift evolution have been 
recognized as sensitive probes of the normalization of the power spectrum
and the mean matter density in the nearby universe 
(White et al. 1993b; Viana \& Liddle 1999).
The parameter degeneracy between $\sigma_8$, the {\it rms} amplitude of mass
fluctuations in the universe filtered on an 8$h^{-1}$~Mpc scale, and the 
matter density parameter $\Omega_m$ can be broken by extending cluster
surveys to higher redshift (Bahcall et al. 1997).  This particular probe
is highly complementary to the CMB anisotropy, because it probes the era of
structure formation-- when dark energy becomes dominant-- as opposed to
the era of recombination.

\begin{figure}[htb]
\hbox to \hsize{
\vbox to 2.25in{\epsfysize=2.65in\epsfbox{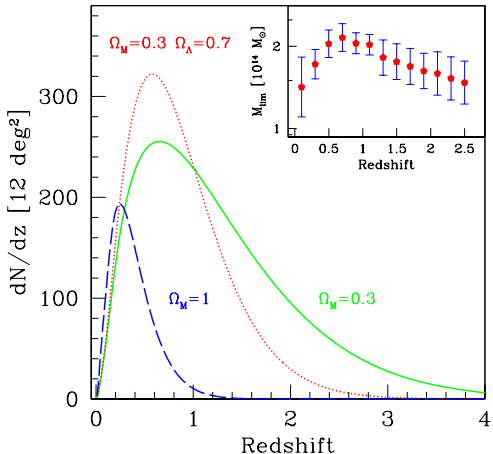}\vfil}\hfil}
\vskip-2.3in
\caption{\hangafter=-16\hangindent=2.40in
The cluster redshift distribution of an interferometric SZE 
survey similar to the one planned with the SZA.  All curves are 
normalized to produce the observed local abundance of massive clusters.
The redshift distribution is sensitive to cosmological parameters.  
Note the inset, which shows the limiting mass of clusters, which 
result in $5\sigma$ detections in mock SZE observations.}
\vskip0.3in
\end{figure}

The observed cluster redshift distribution in a survey (see Figure~2)
is the comoving volume per unit redshift and solid angle 
$dV/dz\,d\Omega$ times
the comoving density of clusters $n_{com}$ with masses above the survey
detection limit $M_{lim}$:
written as
\begin{equation}
{dN\over dz\, d\Omega} = {dV_{com}\over dz\, d\Omega} n_{com}=
{c\over H(z)} d_A^2(z)\left(1+z\right)^2 
\int_{M_{lim}(z)}^{\infty} dM\, {dn\over dM},
\label{eq:dndz}
\end{equation}
where $dn/dM$ is the cluster mass function, $H(z)$ is the Hubble parameter
as a function of redshift and $d_A$ is the angular diameter distance.  The
cosmological sensitivity comes from the three basic elements:
\begin{list}{$\bullet$}{
\setlength{\leftmargin}{\parindent}
}

\item {\bfseries Volume:}  the volume per unit solid angle and 
redshift depends sensitively
on cosmological parameters (i.e. higher $\Omega_\Lambda$ or lower $\Omega_m$
increases the volume per solid angle).  Figure 3 (left) is a 
plot of the comoving volume element ($dV/dz/d\Omega$) versus redshift for three cosmological 
models.  Note the rapid increase in the volume element at modest 
redshift, which is responsible for the rapid rise in the cluster 
redshift distribution in Figure~2.  At higher redshift the comoving volume 
element flattens out and eventually turns over.

The cosmological sensitivity of the distance-redshift and 
volume-redshift relation derives essentially from the expansion 
history of the universe $E(z)$, where $H(z)=H_{0}E(z)$, where $H_{0}$ is the 
Hubble parameter and the parameter $E(z)$ describes its evolution.  
Within our cosmological framework, the expansion history of the 
universe simply depends on the nature and amount of the constituents 
that make up the universe.  That is, 
$E^{2}(z)=\Omega_{M}\left(1+z\right)^{3}+
\left(1-\Omega_{M}-\Omega_{E}\right)\left(1+z\right)^{2}+
\Omega_{E}\left(1+z\right)^{3(1+w)}$.

\item {\bfseries Abundance:} the number density of clusters at a 
given redshift depends sensitively on the growth rate of 
density perturbations.  This growth rate is
highly sensitive to cosmology (i.e. higher $\Omega_m$ speeds the growth
of density perturbations so that clusters ``disappear'' more quickly
as we probe to higher redshift).  Figure~3 (right) is a plot of the 
comoving abundance of clusters above a fixed mass, where the 
abundance is normalized to reproduce the observed local abundance of 
massive clusters.  Note that abundance differences increase 
dramatically with redshift and are responsible for the high redshift 
($z>1$) behavior of the cluster redshift distribution (Figure~2).

As stated above, the cosmological sensitivity of the abundance evolution appears to 
derive from the growth rate of density perturbations.  
Within the linear regime, the differential equation that 
describes growth depends, again, on the expansion history of the 
universe $E(z)$.  The rapid evolution of the abundance is due to an 
exponential dependence of abundance upon the amplitude of density 
fluctuations on the galaxy cluster scale (Press \& Schechter 1974;
Jenkins et al 2001).

\begin{figure}[htb]
\hbox to \hsize{\hfil
\vbox to 2.25in{\epsfysize=2.65in\epsfbox{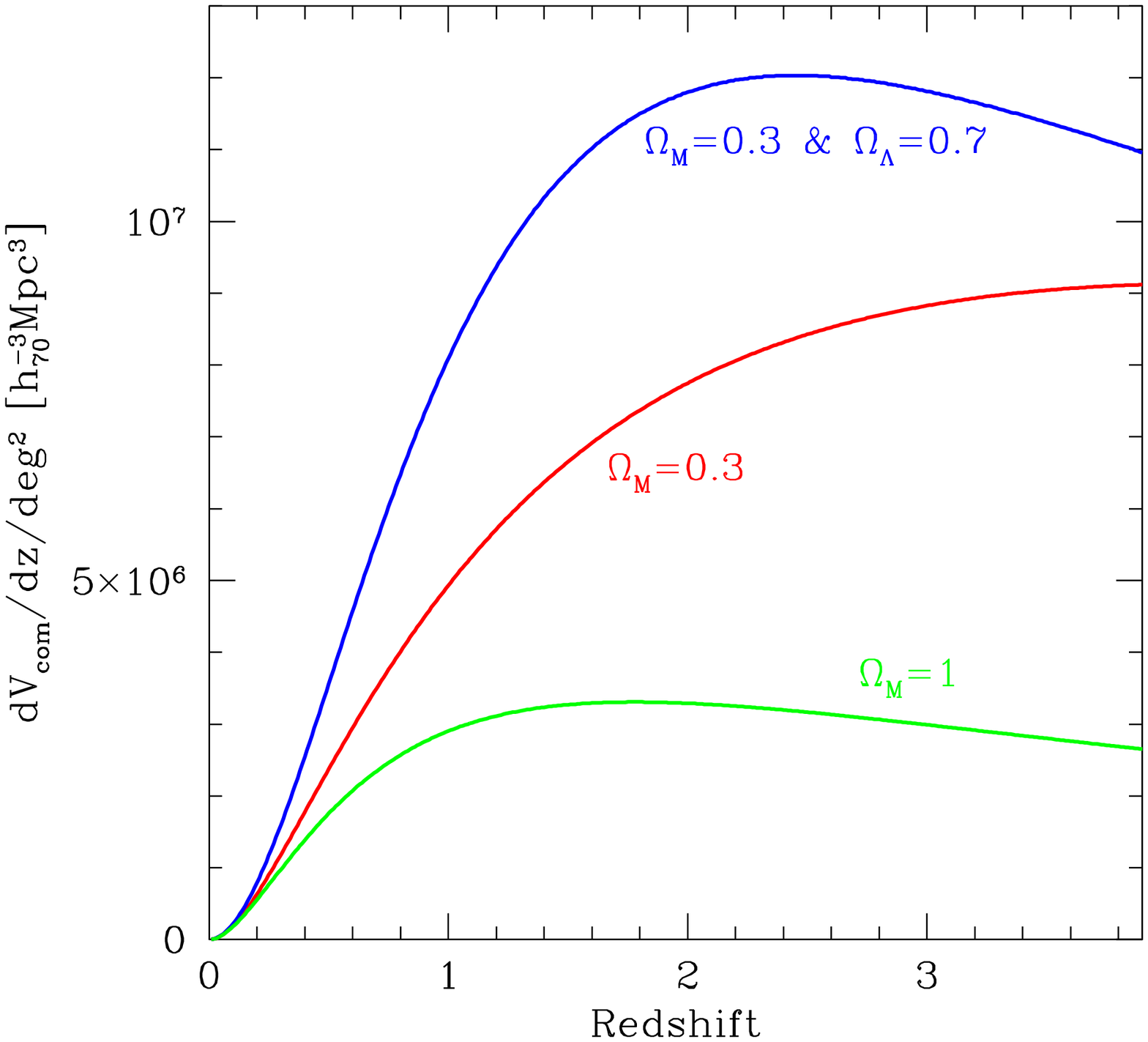}\vfil}
\vbox to 2.25in{\epsfysize=2.65in\epsfbox{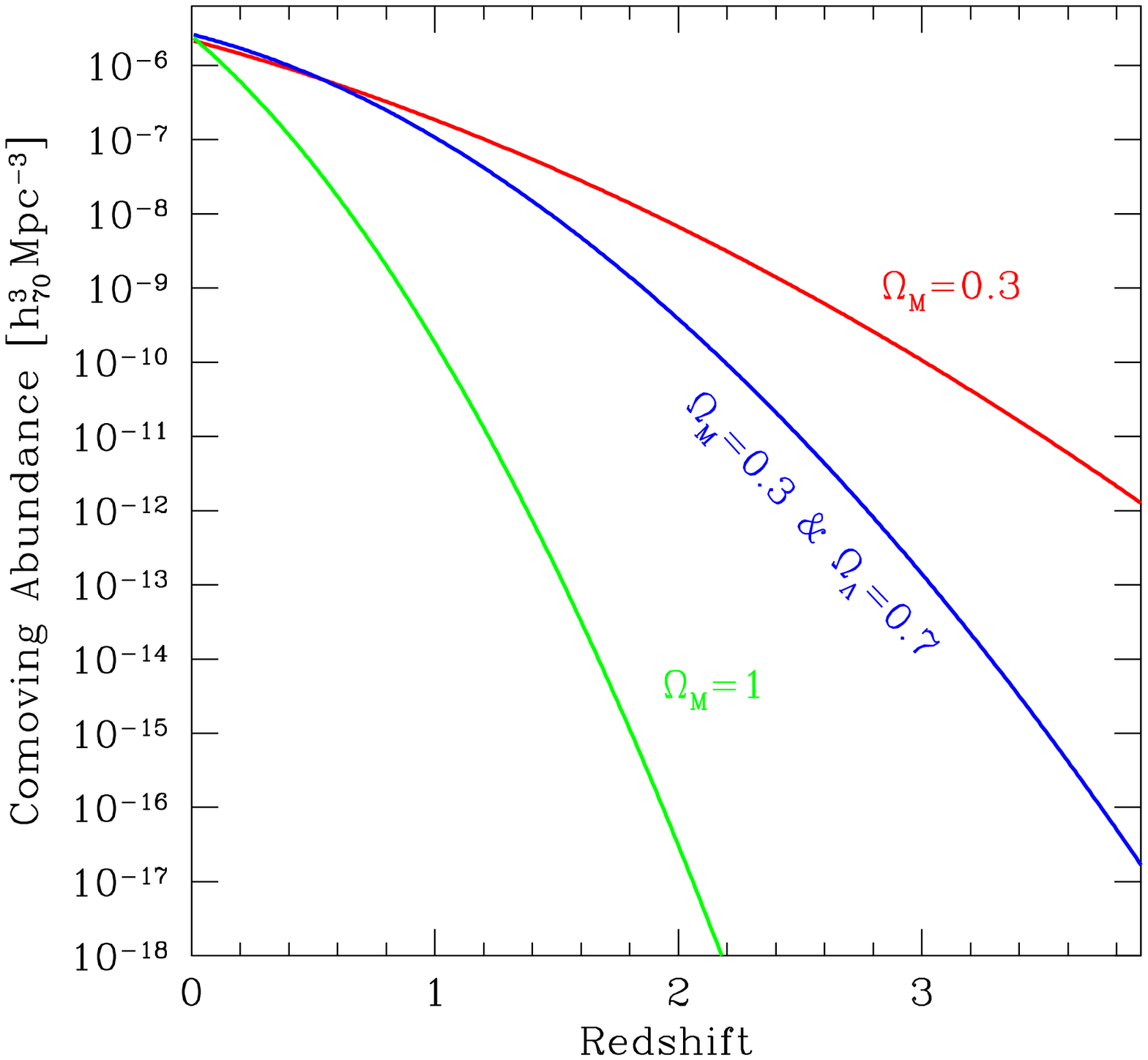}\vfil}
\hfil}
\caption{
The comoving volume element (left) and cluster abundance above 
a fixed mass (right) in three different cosmological models.  The 
abundances are normalized to produce the observed local abundance of 
massive clusters.  Differences in the cluster redshift 
distribution are dominated by volume at low redshift and by 
abundance at high redshift.}
\end{figure}

\item {\bfseries Mass limit:}  the mass of a cluster,
which is just luminous enough to appear above the detection threshold,
typically depends on the luminosity or angular diameter
distance as well as the evolution of cluster structure-- both are sensitive
to cosmological parameters.  The survey yield and redshift 
distributions are both sensitive to the limiting mass, as indicated in 
Figure~4.  Figure~4 shows the cluster redshift distribution in a 
fiducial cosmology for a limiting mass of $M=2\times10^{14}M_{\odot}$, 
and for limiting masses 10\% above and below this value.

\begin{figure}[htb]
\hbox to \hsize{
\vbox to 2.25in{\epsfysize=2.65in\epsfbox{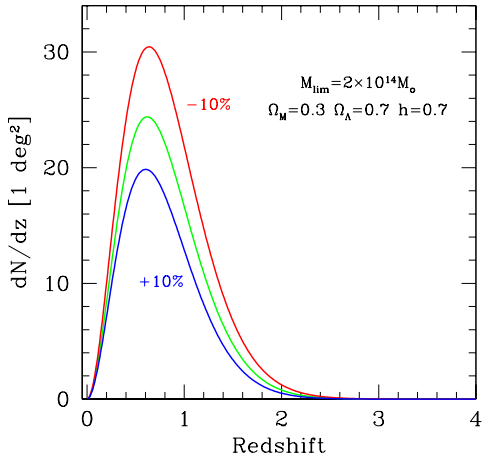}\vfil}\hfil}
\vskip-2.30in
\caption{\hangafter=-16\hangindent=2.40in
The cluster redshift distribution within a fiducial cosmological model
for a mass limit of $M=2\times10^{14}M_{\odot}$ and for mass limits
10\% above and below this value.  The mass sensitivity of the survey 
yields and redshift distributions means that accurate cosmological constraints 
require unbiased estimators of cluster halo mass.}
\vskip0.4in
\end{figure}

\end{list}

The challenging aspects of using galaxy clusters to constrain 
cosmology, aside from building the instruments to carry out the 
surveys, include an understanding of how cluster abundance evolves
within a variety of cosmologies and how to relate cluster observables 
like X-ray emission, SZE distortion, galaxy light and weak lensing 
shear to halo mass.  These relations are required for all redshifts.  
Theoretical studies of structure formation suggest that the mass 
function $dn/dM(z)$ is well behaved, and may be described by a 
``universal'' form when suitably parametrized (Jenkins et al 2001;
White 2001).  Further study is clearly required.  Observational studies
of galaxy cluster scaling relations suggest regularity in the cluster
population similar to the regularity in the elliptical galaxy population
(Mohr \& Evrard 1997; Mohr, Mathiesen \& Evrard 1999; Horner, Mushotzky
\& Scharf 1999).  Hydrodynamical 
simulations of cluster formation suggest that scaling relations 
between cluster observables and halo mass evolve in a simple way
(Evrard et al. 1996; Bryan \& Norman 1998), even in the presence of
some early preheating (Bialek, Evrard \& Mohr 2001).  Much more study using 
higher resolution simulations that incorporate additional physics is 
clearly required here to enable more accurate, unbiased estimators of 
cluster mass.  

A recent study by Diego et al (2001) 
suggests that a joint analysis of
the cluster redshift distribution and the observed 
scaling relations (all available from the same survey data) can 
allow one to solve for the evolving scaling relation and 
cosmological parameters simultaneously (see also Verde, Haiman \& 
Spergel 2001).  More complete studies of the degeneracies between
the evolution of cluster scaling relations and cosmological parameters
in the analysis of cluster surveys is ongoing.

\begin{figure}[thb]
\hbox to \hsize{\hskip0.1in
\vbox to 2.8in{\epsfysize=3.0in\epsfbox{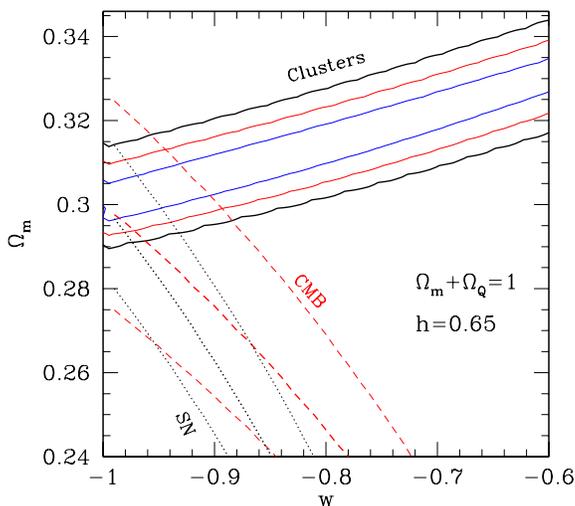}\vfil}\hfil}
\vskip-2.92in
\caption{
\hangafter=-16\hangindent=2.85in
We plot estimated confidence regions for joint constraints on $\Omega_m$ and
the equation of state parameter $w$ of the dark energy from an X--ray survey
proposed in a NASA Small Explorer competition.  All cosmological models
are flat, and a fiducial model of $\Omega_m=0.3$, $\Omega_\Lambda=0.7$ and 
$w=-1$ is assumed.  Also shown are the confidence regions which correspond
to a 1\% measurement of the angular scale of the first acoustic peak
in the CMB anisotropy and a 1\% measurement of the luminosity distance to $z=1$.
This figure illustrates the power of a cluster survey, which yields $\sim10^3$
clusters with measured temperatures.  In addition, it shows that the parameter
degeneracy in the cluster constraints is roughly orthogonal to the parameter
degeneracy from the CMB or SNe Ia measurements.
(figure from Haiman, Mohr \& Holder 2001)}
\label{fig:HMH}
\end{figure}

\section{Precision Cosmology with Galaxy Cluster Surveys?}

Recently, Haiman, Mohr \& Holder (2001) emphasized that large 
cluster surveys extending to
high redshift can in principle provide precision measurements of any
cosmological parameter, which affects the expansion history of the universe
(i.e. $\Omega_m$, $\Omega_\Lambda$, and $w$).  
This is emphasized in Figure~5, which
shows the 1$\sigma$, 2$\sigma$ and 3$\sigma$ joint
constraints on $\Omega_m$ and the equation of state parameter $w$ of 
the dark energy for an X--ray cluster survey which yields $\sim10^3$
clusters with measured emission weighted mean temperatures (and therefore
virial mass estimates).  Only flat models ($\Omega_m+\Omega_\Lambda=1$)
are considered and the fiducial model $\Omega_m=0.3$, $\Omega_\Lambda=0.7$,
constant $w=-1$ and $h=0.65$ is adopted.  
Note the $\Omega_m-w$ degeneracy.  Also shown (dashed line) is
the $\Omega_m-w$ degeneracy for CMB anisotropy and SNe Ia distance
measurements.  The CMB degeneracy assumes that the angular scale of
the first peak (at fixed $h=0.65$) is known to an accuracy of 1\%, whereas
the SNe Ia degeneracy assumes that the luminosity distance to $z=1$ is
known to 1\%.  This figure indicates that an X--ray survey yielding
$10^3$ clusters has comparable constraining power to 1\% CMB or SNe Ia
measurements.  In addition, the roughly orthogonal degeneracy between
the cluster constraints and those from the CMB and SNe emphasizes the
complementarity of these independent constraints on cosmological parameters.

The requirements for such precision are (1) a large cluster sample 
extending to intermediate or high redshift and (2) cluster mass 
estimators that are unbiased at the $\sim$5\% level.  Assuming these two 
requirements can be satisfied, cluster surveys have as much potential 
to reveal the nature and amount of the dark energy in our universe as 
either high redshift type Ia supernovae observations or observations 
of the anisotropy of the cosmic microwave background.  

\begin{figure}[htb]
\hbox to \hsize{
\vbox to 2.25in{\epsfysize=2.65in\epsfbox{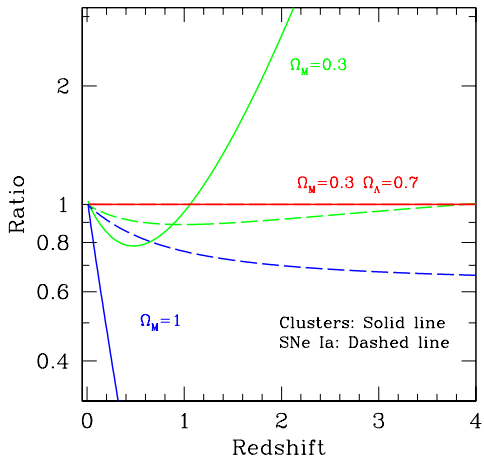}\vfil}\hfil}
\vskip-2.35in
\caption{\hangafter=-16\hangindent=2.40in
The ratio of the cluster redshift distribution (solid line) and the 
luminosity distance (dashed line) as a function of redshift for three 
cosmological models.  The fiducial model is $\Omega_{M}=0.3$ and 
$\Omega_{\Lambda}=0.7$, and the redshift distribution and distances 
from this model appears in the denominator of all ratios.  Aside from 
a narrow window around $z=1$, cluster redshift distributions are far 
more sensitive to cosmological parameters than are luminosity 
distances.}
\end{figure}

In fact, 
with accurate mass estimators, the cluster redshift distribution is 
far more cosmologically informative than simple distance 
measurements.  Figure~6 contains a plot, which compares the 
cosmological sensitivity of SNe Ia distance estimates to cluster 
survey constraints, assuming distances and cluster masses are both 
accurately estimated from the data.  The figure plots the ratio of the 
cluster redshift distribution (solid lines) and the luminosity 
distance (dashed lines) as a function of redshift for a few 
cosmological models.  The denominator in each ratio is the quantity 
from the fiducial model, taken to be $\Omega_{M}=0.3$ 
and $\Omega_{\Lambda}=0.7$ in this example.  The degree to which the ratio 
deviates from 1.0 provides an indication of the sensitivity to 
differences in the two models.  Except for a 
narrow window around $z=1$, cluster redshift distribution contains 
more cosmological information than do luminosity distances.   Given 
the discussion in Section~3 above, this is easy to understand.  Surveys 
probe the volume--redshift relation, which scales as the square of 
the distance.  Abundance evolution depends exponentially on the 
growth rate of density perturbations. In addition, limiting masses 
depend on luminosity or angular diameter distances.  Naturally, these 
different dependencies can interfere constructively or destructively.

Of course, cluster surveys will only achieve high precision if 
cluster masses can be accurately estimated, on the average, from observables.  
That is, precision requires that systematic biases in mass estimators be small.
This is similar to the case with type Ia supernovae, in that SN Ia distances 
are only accurate to the extent that the SNe themselves are 
standard candles.  The high potential of both approaches has 
led people to invest significant effort in better understanding 
possible sources of systematics.  Currently, our theoretical 
understanding of the formation and evolution of clusters is
less developed than our understanding of the dynamics of density 
perturbations (well within the linear regime) at and before the epoch of 
recombination; however, it seems to me that our understanding 
of structure formation has progressed well beyond our understanding 
of why SNe Ia's form a one parameter family of standard candles that 
have not evolved since before the universe was one quarter its present age.
This theoretical heritage in structure formation is an important resource
as we move toward interpreting ongoing and planned surveys.

Precision cosmology with clusters requires large cluster ensembles 
extending over large ($10^{2}$-$10^{3}$~deg$^{2}$) solid angles.  
X--ray and SZE surveys of this sort will not necessarily have the 
optical/near-IR data available on every system to estimate a 
sufficiently accurate redshift.  Nevertheless, X--ray and SZE
surveys are attractive, because high signal to noise detections are possible,
projection effects are minimized, and observations indicate that tight scaling
relations involving X--ray observables exist.
The price of not having redshifts can 
be severe, as shown in Figure~7.  This figure shows the confidence 
regions in $\Omega_{M}$ and $w$ corresponding to a 4000~deg$^{2}$ SZE  observations indicate that tight scaling
relations involving X--ray observables exist.
survey carried out from the South Pole.  Confidence regions include 
marginalization over $\sigma_{8}$, and only flat models are 
considered.  The contours correspond to constraints using only the 
total number of detected clusters, whereas the solid region denotes 
the constraints in the case that cluster redshifts are available.  As 
was made clear in Section~3, the cluster redshift distribution
is cosmologically rich.

Possibilities for redshift followup include large solid angle, 
multiband optical or near-IR photometric surveys and direct 
spectroscopic followup of member galaxies. It may even be possible to  
obtain rough redshift estimates directly from the SZE or X--ray data, 
but further work is required to explore the feasibility of this 
approach (Diego et al, in preparation).
Photometric redshift estimates of multiple galaxies within each 
cluster should be sufficiently accurate to allow precision 
cosmology.  The required redshift precision is set less by the scale 
of change in the theoretical cluster redshift distributions (which are 
smoothly varying- see Figure~2) than
by the fact that accurately inferring cluster masses 
from measured fluxes requires redshifts.

\begin{figure}[h]
\hbox to \hsize{
\vbox to 2.25in{\epsfysize=2.65in\epsfbox{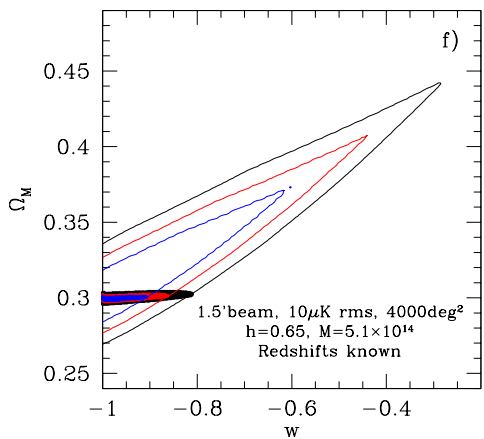}\vfil}\hfil}
\vskip-2.35in
\caption{\hangafter=-16\hangindent=2.40in
Confidence regions on $\Omega_{M}$ and mean $w$ from a bolometric SZE 
survey carried out over 4000~deg$^{2}$ with a proposed South Pole 8~m 
telescope.  This survey yields $\sim$20,000 clusters extending to 
redshifts $z=2$.  This plot underscores the importance of
having redshifts for detected clusters;  the solid confidence 
regions correspond to 
constraints from the cluster redshift distribution, whereas the 
contours correspond to confidence regions from the total number of 
detected clusters.}
\end{figure}

The Sloan Digital Sky Survey is one such multiband, photometric 
dataset, and it will be extremely useful in 
estimating redshifts for clusters at $z<0.6$ (i.e. relatively low 
redshift samples like the high mass {\it Planck} Surveyor cluster 
sample).  However, for reasonably deep X--ray surveys and high 
sensitivity SZE surveys, only a small fraction of the sample will lie 
at these low redshifts, and so deeper, multiband followup will be 
required.  The bad news is that the effort required to carry out 
deep, multiband surveys in the optical and near-IR is comparable 
to the effort required to execute the initial SZE or X-ray survey.  The 
good news is that there are several projects being designed 
independently of planned and proposed SZE and X-ray surveys that will 
provide the required data.  These survey projects include PRIME--- a 
NASA Small Explorer Class proposal in Phase A study that will survey 
one quarter of the sky in the 1-3$\mu$m range, VISTA--- a 4~m 
class telescope with a wide field near-IR and (eventually) optical 
camera that will carry out surveys in the southern hemisphere,  the 
VST--- an SDSS-like survey telescope with a large optical camera 
operating in the southern hemisphere, and the Large Synoptic 
Survey Telescope--- a 6--8~m class telescope with wide field of view 
to carry out frequent, repeated imaging of large portions of the 
sky.  In addition, there are extremely useful large field of view CCD 
cameras available at KPNO/CTIO and on the CFHT.  With these 
projects and others together with the exciting science possible with 
cluster surveys, it is only a question of time until very large 
cluster catalogs can be derived from large solid angle,
multifrequency surveys.

\section{Discussion}

This contribution contains a description of two ways of using cluster 
surveys to learn about structure formation and cosmology: (1) the 
context free test of hierarchichal structure formation using SZE 
cluster surveys, which are sufficiently sensitive to detect low mass 
clusters no matter what their redshift, and (2) the use of cluster 
redshift distributions within the context of our standard model for 
structure formation to determine the quantity and nature of dark 
matter and dark energy in the universe.  It's important to emphasize 
that there is additional information that comes with a cluster 
survey.  This information allows one to study the cluster mass 
function $dn/dM$ as a function of redshift, likely improving the 
constraints derived from integrals over the mass function (i.e. 
equation~3).   In addition, surveys (perhaps with some targeted 
followup) enable one to study cluster scaling 
relations such as the X-ray, optical or SZE luminosity--temperature 
or luminosity--mass relations;  a combined study of scaling relations 
and the redshift distribution may well allow one to solve for the 
scaling relation evolution and cosmological parameters simultaneously 
(Diego et al 2001).  

One can also study the spatial correlations 
among clusters to infer properties of the underlying power spectrum 
of dark matter density fluctuations.  With good halo mass estimates 
like those required to use the cluster redshift distribution to full 
effect, it should be possible to use the cluster power spectrum 
constraints to improve limits on the neutrino mass density.  
Even in the absence 
of accurate halo mass estimates, it should be possible to use large 
surveys (in volume and number) to measure the scale of the break in 
the transfer function for the evolution of density perturbations. 
Recently, Cooray et al (2001) have emphasized that
the physical scale of the break in the transfer function, which is 
the horizon scale at matter--radiation equality, depends on the matter 
density and CMB temperature.  The matter density is measured to high 
precision with CMB anisotropy observations such as those with MAP and 
Planck.  Therefore, the break in the transfer function is a standard 
rod, whose scale is independent of redshift and is calibrated to high 
accuracy with CMB data.  Thus, measurements of the cluster 
correlation function within redshift shells returns the 
angular diameter distance as a function of redshift, much like the 
SNe Ia but with a strong physical basis for the lack of evolution in 
the standard rod.  This approach is very complementary to the cluster 
redshift distribution approach, and it hinges less on extracting 
unbiased estimates of cluster masses from cluster observables like the 
X-ray or SZE luminosity. 

\acknowledgements
I would like to acknowledge Zoltan Haiman for the many cluster 
survey discussions and calculations we have shared.  It is also a 
pleasure to acknowledge many cluster related conversations with 
John Carlstrom and Gil Holder.

\end{document}